\documentclass[superscriptaddress,secnumarabic,nobibnotes,aps,prd,showkeys,showpacs,nofootinbib,onecolumn,eqsecnum]{revtex4}
\usepackage{graphics}
\usepackage{graphicx}
\usepackage{color}
\usepackage{epsf}
\usepackage{bm}
\usepackage{amsmath,amssymb,amsfonts,mathrsfs,amsthm}
\usepackage{latexsym}
\usepackage{enumerate}
\usepackage{comment}
\usepackage{hyperref}
\usepackage{amsmath}
\usepackage{amsfonts}
\usepackage{amssymb}

\providecommand{\U}[1]{\protect\rule{.1in}{.1in}}

\DeclareMathOperator{\arcsinh}{arcsinh}

\newcommand{\be}{\begin{equation}}
\newcommand{\ee}{\end{equation}}
\newcommand{\bea}{\begin{eqnarray}}
\newcommand{\ea}{\end{eqnarray}}
\newcommand{\ben}{\begin{equation*}}
\newcommand{\een}{\end{equation*}}
\newcommand{\bean}{\begin{eqnarray*}}
\newcommand{\eean}{\end{eqnarray*}}
\def\bal#1\eal{\begin{align}#1\end{align}}

\newcommand{\mincir}{\raise
-3.truept\hbox{\rlap{\hbox{$\sim$}}\raise4.truept\hbox{$<$}\ }}
\newcommand{\magcir}{\raise
-3.truept\hbox{\rlap{\hbox{$\sim$}}\raise4.truept\hbox{$>$}\ }}

\begin{document}

\title{Quantization of the Szekeres System}
\author{A. Paliathanasis}
\email{anpaliat@phys.uoa.gr}
\affiliation{Instituto de Ciencias F\'{\i}sicas y Matem\'{a}ticas, Universidad Austral de
Chile, Valdivia, Chile}
\affiliation{Institute of Systems Science, Durban University of Technology, POB 1334
Durban 4000, South Africa.}
\author{Adamantia Zampeli}
\email{azampeli@phys.uoa.gr}
\affiliation{Nuclear and Particle Physics section, Physics Department, University of
Athens, 15771 Athens, Greece}
\affiliation{Institute of Theoretical Physics, Faculty of Mathematics and Physics,
Charles University, V Hole\v{s}ovi\v{c}k\'ach 2, 18000 Prague 8, Czech
Republic}
\author{T. Christodoulakis}
\email{tchris@phys.uoa.gr}
\affiliation{Nuclear and Particle Physics section, Physics Department, University of
Athens, 15771 Athens, Greece}
\author{M.T. Mustafa}
\email{tahir.mustafa@qu.edu.qa}
\affiliation{Department of Mathematics, Statistics and Physics, College of Arts and
Sciences, Qatar University, Doha 2713, Qatar}

\begin{abstract}
{We study the quantum corrections on the Szekeres system in the context of
canonical quantization in the presence of symmetries. We start from an
effective point-like Lagrangian with two integrals of motion, one
corresponding to the Hamiltonian and the other to a second rank Killing
tensor. Imposing their quantum version on the wave function results to a
solution which is then interpreted in the context of Bohmian mechanics. In
this semiclassical approach, it is shown that there is no quantum
corrections, thus the classical trajectories of the Szekeres system are not
affected at this level. Finally, we define a probability function which
shows that a stationary surface of the probability corresponds to a
classical exact solution.}
\end{abstract}

\keywords{Szekeres system; Silent universe; Quantization; Semiclassical
approach}
\maketitle
\date{\today}

%-----------------------------------------------------------------------------------------------------------------------------------------------------------------------------------------------------------------------------------------------

%-----------------------------------------------------------------------------------------------------------------------------------------------------------------------------------------------------------------------------------------------

%-----------------------------------------------------------------------------------------------------------------------------------------------------------------------------------------------------------------------------------------------
%-----------------------------------------------------------------------------------------------------------------------------------------------------------------------------------------------------------------------------------------------
%-----------------------------------------------------------------------------------------------------------------------------------------------------------------------------------------------------------------------------------------------

\section{Introduction}

%-----------------------------------------------------------------------------------------------------------------------------------------------------------------------------------------------------------------------------------------------
{The silent universe is mathematically described by a set of six first-order
differential equations following from the assumptions that i) the magnetic
part of the Weyl tensor vanishes and ii) that the total matter source of the
universe is described by a pressureless perfect fluid (irrotational dust
fluid component) \cite{silent1,silent2,silent3}. In physical terms, the main
property of the silent universe is that there is no information
dissemination through gravitational or sound waves. }

{It is known that there exists a family of exact solutions for the field
equations of the silent universe described as Szekeres geometries with line
element of the form \cite{szek0}
\begin{equation}
ds^{2}=-dt^{2}+e^{2\alpha}dr^{2}+e^{2\beta}\left( dy^{2}+dz^{2}\right)
\label{ss.04aa}
\end{equation}
where $\alpha \equiv \alpha\left( t,r,y,z\right) $ and $\beta
\equiv\beta\left( t,r,y,z\right)$. There exists two families of solutions
which correspond to the Friedmann--Lema\^{\i}tre--Robertson--Walker (like)
geometries and the Kantowski-Sachs solutions \cite{kraj}. One of the
simplest solution in the Szekeres family is the Bondi-Tolman metric with
line element
\begin{equation*}
ds^{2}=-dt^{2}+\left( t-t_{0}\left( r\right) \right) ^{-\frac{2}{3}}\left(
t-t_{0}\left( r\right) +\frac{2}{3}r\frac{dt_{0}\left( r\right) }{dr}\right)
^{2}dr^{2}+2\frac{\left( t-t_{0}\left( r\right) \right) ^{\frac{4}{3}}}{%
\left( 1+y^{2}+z^{2}\right) ^{2}}\left( dy^{2}+dz^{2}\right)
\end{equation*}
from which it is clear that the position of the singularity $t=t_{0}\left(
r\right)$ is space dependent, for more details see \cite{kraj}. }

{In these solutions, the two components of the electric part of the Weyl
tensor and the two components of the shear for the observer $u^{\mu}$ are
equal respectively. While someone would expect that the ``symmetry" between
the different components of the Weyl tensor and the shear will generate an
extra Killing field in the underlying manifold, it was shown in \cite{musta}
that the spacetimes coming from the Szekeres system are actually
``partially" locally rotationally symmetric and not exactly. The particular
interest of the scientific community on these geometries lies on their
interesting properties and on the fact that they can be seen as
inhomogeneous models \cite{sm00,sm00a0,sm00a,sm01}; this property renders
these spacetimes proper for the description of FLRW spacetimes perturbations
\cite{silent1}. Some recent results on the Szekeres geometries with one
isometry and on their conformal symmetries can be found in \cite{ks1,ks2}. }

{In the following, we consider a Riemannian manifold with metric $g_{\mu\nu}$
and a timelike four-vector field $u^{a}$. Let $T_{\mu\nu}$ be the energy
momentum tensor of the matter source; the energy density is defined as $%
\rho=T^{\mu\nu }u_{\mu}u_{\nu}$ and the field equations of the silent
universe are reduced to a system of algebraic-differential equations with
the algebraic equation being
\begin{equation}
\frac{\theta^{2}}{3}-3\sigma^{2}+\frac{^{\left( 3\right) }R}{2}=\rho,
\end{equation}
and the first-order differential equations
\begin{subequations}
\label{szeksys}
\begin{align}
&\dot{\rho}+\theta\rho =0,~  \label{ss.01} \\
&\dot{\theta}+\frac{\theta^{2}}{3}+6\sigma^{2}+\frac{1}{2}\rho =0,
\label{ss.02} \\
&\dot{\sigma}-\sigma^{2}+\frac{2}{3}\theta\sigma+E =0,  \label{ss.03} \\
&\dot{E}+3E\sigma+\theta E+\frac{1}{2}\rho\sigma =0,  \label{ss.04}
\end{align}
where $\dot{}$ denotes the directional derivative along $u^{\mu}$, i.e. $%
\dot{}=u^{\mu}\nabla_{\mu}$. The set of equations \eqref{szeksys} is also
well-known as Szekeres system \cite{szek0,barn1}. The parameter $\theta$ is
the expansion rate of the observer, $\theta=\left(
\nabla_{\nu}u_{\mu}\right) h^{\mu\nu}$, while $\sigma~$and $E$ are the shear
and electric component of the Weyl tensor, $E_{\nu}^{\mu}=$ $%
Ee_{\nu}^{\mu},~\sigma_{\nu}^{\mu }=\sigma e_{\nu}^{\mu},$ in which the set
of $\left\{ u^{\mu} ,e_{\nu}^{\mu}\right\} $ defines an orthogonal tetrad
such that $u_{\mu}e_{\nu}^{\mu}=0; \ e_{\nu}^{\mu
}e_{\mu}^{\lambda}=\delta_{\nu}^{\mu}+u^{\mu}u_{\nu},$ such that the
components of tensors are scalar functions \cite{silent1}. The relation of
the kinematical parameters $\theta,\sigma$ and $E$ to the functions $%
\alpha,\beta$ can be recovered by considering the $3+1$ decomposition of the
line element \eqref{ss.04aa} for an observer $u^{\mu}=\delta_{t}^{\mu}$.
Then, the expansion rate and the shear are \cite{carge1}
\end{subequations}
\begin{equation}
\theta=\left( \frac{\partial\alpha}{\partial t}\right) +2\left( \frac{%
\partial\beta}{\partial t}\right) ~~,~\sigma^{2}=\frac{2}{3}\left( \left(
\frac{\partial\alpha}{\partial t}\right) -\left( \frac{\partial \beta}{%
\partial t}\right) \right) ^{2}.
\end{equation}
It is important to note that the complete set of the gravitational field
equations includes the differential equations $h_{\mu}^{\nu}\sigma_{\nu;%
\alpha}^{\alpha}=\frac{2}{3}h_{\mu}^{\nu}\theta_{;\nu}, h_{\mu}^{\nu}
E_{\nu;\alpha}^{\alpha}=\frac{1}{3}h_\mu^{\nu}\rho_{;\nu}$ in which $%
h_{\mu\nu}$ is the decomposable tensor defined by the expression $%
h_{\mu\nu}=g_{\mu\nu}-\frac{1}{u_\lambda u^{\lambda}}u_{\mu}u_{\nu}$ \cite%
{lesame}. Thus, in general, the Szekeres system is a set of partial
differential equations, except when the latter equations are satisfied
identically. In this case, it reduces to a system of ordinary differential
equations.}

Recently, the conservation laws of the Szekeres system \eqref{szeksys} were
constructed with various methods in \cite{anszek} and \cite{szek}. In
particular, in \cite{szek} the method of Darboux polynomials and the Jacobi
multiplier method were applied, while in \cite{anszek} the symmetries and
the movable singularities of the Szekeres system were studied. The novelty
in the analysis of \cite{anszek} is that an effective classical Lagrangian
describing the system \eqref{szeksys} was constructed. Furthermore, it was
shown that the conservation laws of the Szekeres system follow from the
application of Noether's theorem on the aforementioned effective Lagrangian.

{This work explores the effective Lagrangian of \cite{anszek} at the quantum
level by considering its canonical quantization with the use of symmetries
\cite{chris2013}. The aim is to derive the physical properties after
quantization in the context of Bohmian mechanics \cite{boh1,boh2}. The
Bohmian approach to quantum theory is well suited for quantum cosmology,
since it does not presupposes the existence of a classical domain as it is
necessary for the Copenhagen interpretation for the measurement process to
be defined. In addition, this interpretation results in the definition of
deterministic trajectories on the configuration space. This allows the
comparison of the classical versus semiclassical trajectories through the
corresponding properties of each geometry. Hence, its application in
cosmology has been considered before, see e.g. \cite%
{chris2013,pinto,manto,on1,palan}. }

%-----------------------------------------------------------------------------------------------------------------------------------------------------------------------------------------------------------------------------------------------

\section{Classical Dynamics}

%-----------------------------------------------------------------------------------------------------------------------------------------------------------------------------------------------------------------------------------------------
In \cite{anszek} the Szekeres system \eqref{szeksys} was written in an
equivalent form of a two second-order differential equations system
\begin{subequations}
\label{szeksysred}
\begin{align}
&\ddot{x}+2\frac{\dot{y}}{y}\dot{x}-\frac{3}{y^{3}}x=0,  \label{ss.10} \\
&\ddot{y}+\frac{1}{y^{2}}=0.  \label{ss.11}
\end{align}
where the variables $x,y$ are related to the energy density and the electric
term as follows
\end{subequations}
\begin{equation}
\rho=\frac{6}{\left( 1-x\right) y^{3}}, \ E=\frac{x}{y^{3}\left( x-1\right) }%
,  \label{ss.09}
\end{equation}
while the expansion rate and the shear are defined by the equations %
\eqref{ss.01}, \eqref{ss.04} as $\theta=-\frac{\dot{\rho}}{\rho},\sigma=%
\frac{2\left( \dot{\rho}E\mathcal{-}\rho\dot{E}\right) }{\rho\left(
\rho+6E\right) }$. It was shown there that the dynamical system %
\eqref{szeksysred} can be derived by a variational principle with Lagrange
function \cite{anszek}
\begin{equation}
L\left( x,\dot{x},y,\dot{y}\right) =y\dot{x}\dot{y}+x\dot{y}^{2}-xy^{-1}
\label{lan.01}
\end{equation}
This Lagrangian is point-like and describes the motion of a particle in a
two-dimensional space with line element
\begin{equation}
ds_{\gamma}=2\left( ydxdy+xdy^{2}\right)
\end{equation}
under the action of the effective potential $V_{eff}\left( x,y\right)
=xy^{-1}.$ The system \eqref{szeksysred} admits two integrals of motion,
quadratic in the velocities; the first is the Hamiltonian function
\begin{equation}
y\dot{x}\dot{y}+x\dot{y}^{2}-xy^{-1}=h,
\end{equation}
since the system is autonomous, while the second one is the quadratic
function
\begin{equation}
I_{0}=\dot{y}^{2}-2y^{-1},
\end{equation}
which can be constructed by the application of Noether's theorem for contact
symmetries \cite{anszek}.

In order to proceed to the canonical quantization in the next section, we
turn to the Hamiltonian formulation. The canonical momenta are defined by
the Lagrangian \eqref{lan.01} as
\begin{equation}
y^{2}\dot{x}=yp_{y}-2xp_{x}~,~y\dot{y}=p_{x}.  \label{ss.17}
\end{equation}
Then our conserved quantities are written in terms of the momenta
correspondingly as %\begin{subequations}
\begin{align}
&H\equiv\frac{p_{x}p_{y}}{y}-\frac{x}{y^{2}}\left( p_{x}\right) ^{2}+\frac {x%
}{y}=h,  \label{ss.16} \\
&I_{0}=y^{-2}\left( p_{x}\right) ^{2}-2y^{-1}.  \label{sss.17}
\end{align}

\section{Quantization and semiclassical analysis}

%-----------------------------------------------------------------------------------------------------------------------------------------------------------------------------------------------------------------------------------------------
The quantization is based on the idea of promoting the integrals of motion
to operators, thus resulting to two eigenvalue equations. For simplicity, we
choose to work in a new set of variables $\left\{ u,v\right\} $ defined by $%
x=vu^{-1},y=u$, in which the point-like Lagrangian \eqref{lan.01} takes the
form
\begin{equation}
L\left( u,\dot{u},v,\dot{v}\right) =\dot{u}\dot{v}-\frac{v}{u^{2}}.
\label{sss.18}
\end{equation}%
The equations of motion \eqref{szeksysred} become
\begin{subequations}
\begin{align}
& \ddot{u}+u^{-2}=0,  \label{sss.19} \\
& \ddot{v}-2vu^{-3}=0,
\end{align}%
while the quadratic conserved quantity is written as $I_{0}=\dot{u}%
^{2}-2u^{-1}$. Hence, the Hamiltonian and the conserved quantity $I_{0}$ can
be written in terms of the momenta as
\end{subequations}
\begin{subequations}
\begin{align}
& p_{u}p_{v}+\frac{v}{u^{2}}=h,  \label{sss.30} \\
& p_{v}^{2}-2u^{-1}=I_{0}  \label{cons0}
\end{align}%
The canonical quantization proceeds by promoting the Poisson brackets to
commutators, $\{\ ,\ \}\rightarrow \lbrack \ ,\ ]$ and the variables on the
phase space of $(u,v,p_{u},p_{v})$ to operators according to $%
x^{i}\rightarrow \hat{x}^{i}=x^{i},\ p_{i}\rightarrow \hat{p}_{i}=i\frac{%
\partial }{\partial x^{i}}$. This procedure leads to the time-independent
Schr\"{o}dinger equation\footnote{%
In which, $p_{u}p_{v}=-\partial _{uv}$ denotes the Laplace operator $\Box $.
}
\end{subequations}
\begin{subequations}
\label{sss.300}
\begin{equation}
\left( -\partial _{uv}+\frac{v}{u^{2}}\right) \Psi =h\Psi ,  \label{sss.31}
\end{equation}%
and the additional equation
\begin{equation}
\left( \partial _{vv}+\frac{2}{u}\right) \Psi =-I_{0}\Psi ,  \label{sss.32}
\end{equation}%
which follows from the quantization of \eqref{cons0} \cite{ref1,ref2,ref3}.
Contrary to the usual method applied in the literature, we used generalized
symmetries, instead of point symmetries.

The set of equations \eqref{sss.300} provides, through the integrability
conditions which must be satisfied for the consistency of the system, the
following general solution for the wave function
\end{subequations}
\begin{equation}
\Psi \left( I_{0},u,v\right) =\frac{\sqrt{u}}{\sqrt{2+I_{0}u}}\left( \Psi
_{1}\cos f\left( u,v\right) +\Psi _{2}\sin f\left( u,v\right) \right)
\label{generalsol}
\end{equation}%
where~%
\begin{align}
& f\left( u,v\right) =\frac{(hu+I_{0}v)\sqrt{2I_{0}+I_{0}^{2}u}-2h\sqrt{u}%
\arcsinh\sqrt{\frac{I_{0}{u}}{2}}}{I_{0}^{3/2}\sqrt{u}},\quad \text{\ for }%
I_{0}\neq 0,  \label{sss.33} \\
&  \notag \\
& f\left( u,v\right) =\frac{\sqrt{2}\left( hu^{2}+3v\right) }{3\sqrt{u}}%
,\quad \text{for }I_{0}=0.  \label{sss.344}
\end{align}%
The coefficients $\Psi _{1}$ and $\Psi _{2}$ are constants of integration.
It is important to note that, due to the linearity of \eqref{sss.31}, the
general solution is the sum of the expression \eqref{generalsol} on all
possible values of the constant $I_{0}$; that is, $\Psi _{Sol}\left(
u,v\right) =\sum_{I_{0}}\Psi \left( I_{0},u,v\right) $.
%----------------------------------------------------------------------------------------------------------------------------------------

\subsection{Semiclassical analysis}

%----------------------------------------------------------------------------------------------------------------------------------------
In order to find the quantum effect on the classical system, we follow the
Bohmian interpretation of quantum theory \cite{boh1,boh2}. In this context,
the departure from the classical theory is determined by an additional term
in the classical Hamilton-Jacobi equation,
\begin{equation}
\frac{1}{2}G^{\mu \nu }\partial _{\mu }S\partial _{\nu }S+V\left( u,v\right)
+h+Q_{V}\left( u,v\right) =0  \label{sss.34}
\end{equation}%
known as quantum potential and defined by
\begin{equation}
Q_{V}=-\frac{\Box \Omega }{2\Omega }.
\end{equation}%
$\Omega $ denotes the \textbf{amplitude of the wave function in polar form},
$\Psi (u,v)=\Omega (u,v)e^{iS(u,v)}$ and $\Box $ the Laplacian operator of %
\eqref{sss.31}, see e.g. \cite{kim,pinto,manto,Stu2,sing1,palan} and
references therein.

{When the quantum potential is zero, the identification
\begin{equation}  \label{semiclas}
\frac{\partial S}{\partial q_i}=p_{i}=\frac{\partial }{\partial\dot{q}_{i}}
\end{equation}
is possible since the equation \eqref{sss.34} becomes the classical
Hamilton-Jacobi equation; these of course should give the classical solution
of the Euler-Lagrange equations. If this classical definition for the
momenta is retained even when $Q\neq0$, we can find semiclassical solutions
which will differ from the classical ones. }

{\
%---------------------------------------------------------------------------------------------
In our case, we assume that the quantum corrections in the general solution %
\eqref{generalsol} follow from the ``frequency $I_{0}$" with the highest
peak in the wave function. This is in agreement with the so-called Hartle
criterion \cite{hartle}; at the same time, $I_{0}$ is the classical
observable value, since it is an integration constant for the equations of
motion. The assumption that the wave function has survival oscillatory term
leads to the result that, in all possible cases, i.e. for $I_0 \neq 0$ or $%
I_0 = 0$, as well as the subcases $h\neq 0$ or $h=0$, the quantum potential
vanishes, thus providing no quantum corrections. Indeed, solving the set of
the corresponding semiclassical equations \eqref{semiclas} for our
variables, we find the classical solution, since the phase function $S$
which now comes from the wave function is constant. }

{Hence, the wave function which followed from canonical quantization of the
effective Lagrangian \eqref{sss.18} indicates that the Szekeres universe
remains \textquotedblleft silent\textquotedblright , even at the quantum
level.} \textbf{That means that because the quantum potential is zero, i.e. }%
$Q=0$\textbf{, the original system (\ref{ss.01})-(\ref{ss.04}) remain the
same. Hence, the classical solution corresponds to a silent universe.   }

\subsection{Probability density}

\begin{figure}[tbp]
\includegraphics[height=5cm]{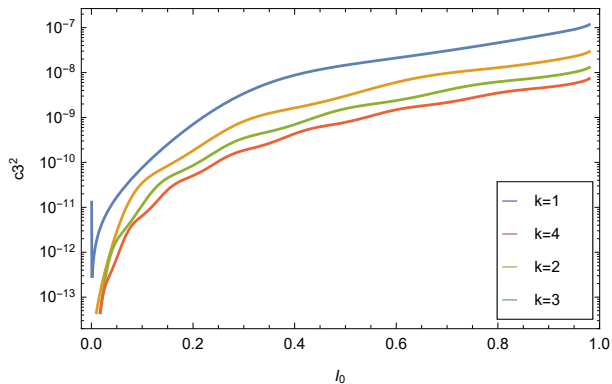}\centering
\caption{Qualitative evolution of the normalize parameter $c_{3}^{2}$ in
terms of the free parameter $I_{0}~$for $k=1$ (blue line), $k=2$, (yellow
line), $k=3$ (green line) and $k=4$ (red line). From the plot we observe
that $c_{3}^{2}$ goes to zero for values of $I_{0}$ close to zero.}
\label{plot111a}
\end{figure}

\begin{figure}[ptb]
\includegraphics[height=6cm]{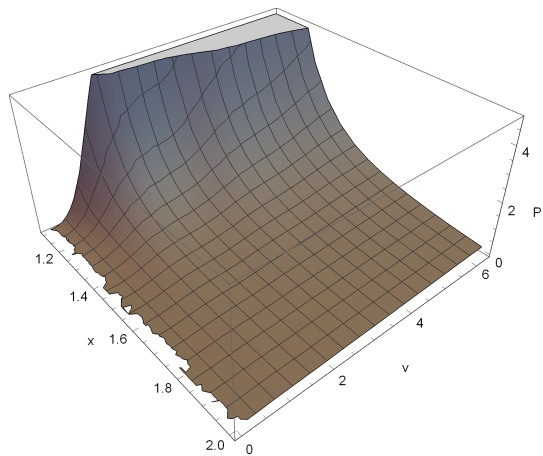}\centering
\caption{Qualitative evolution of the probability function \eqref{probfun}
in the space of variables ${x,v}$. }
\label{plot222}
\end{figure}

\begin{figure}[ptb]
\includegraphics[height=6cm]{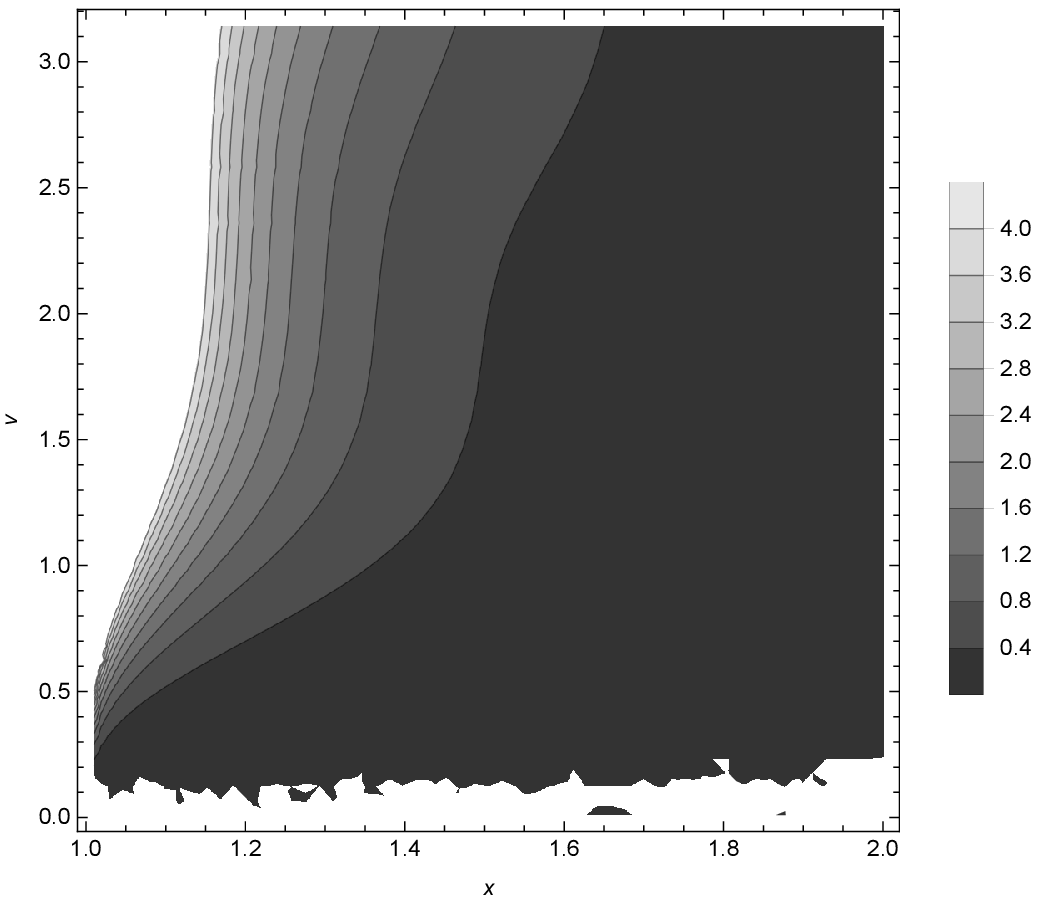}\centering
\includegraphics[height=6cm]{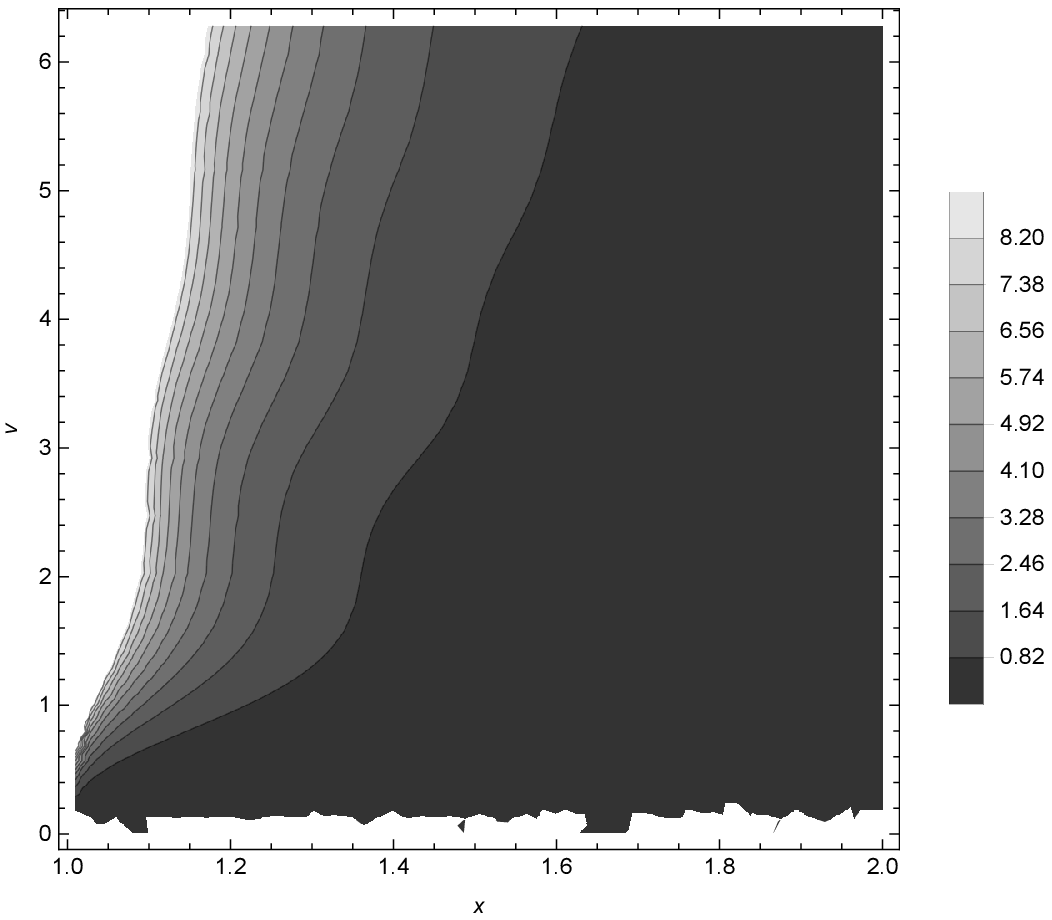}\centering
\caption{Contour plot for the probability function \eqref{probfun} in the
space of variables ${x,v}$. We observe that as $x\rightarrow0$ and $v$ is
small, that is, $I_{0}\rightarrow0$, the function $P\left( x,v\right) $
reaches to a minimum extreme.}
\label{plot333}
\end{figure}

%-----------------------------------------------------------------------------------------------------------------------------------------------------------------------------------------------------------------------------------------------
In this section we restrict ourselves to the case $h=0$ in which the wave
function \eqref{generalsol} with $f(u,v)$ given by \eqref{sss.33} and $h=0$
becomes
\begin{equation}
\Psi _{0}\left( I_{0},u,v\right) =\frac{\sqrt{u}}{\sqrt{2+I_{0}u}}\left(
\Psi _{1}\cos \left( \sqrt{\frac{2+I_{0}u}{u}}v\right) +\Psi _{2}\sin \left(
\sqrt{\frac{2+I_{0}u}{u}}v\right) \right) .
\end{equation}%
The case $\Psi _{1}\rightarrow 0$ which is well behaved at the limits $%
u\rightarrow 0$ and $u\rightarrow \infty $ leads to the following
probability
\begin{equation}
P=\int_{0}^{\infty }du\ dv\mu (u,v)\Psi _{0}^{\ast }\Psi _{0}
\end{equation}%
where $\mu (u,v)=\sqrt{det{G_{\alpha \beta }}}=1$ is the measure on the
space of the configuration variables $(u,v)$. After a change of the variable
$u\rightarrow \frac{2}{x^{2}-I_{0}}$ which induces the Jacobian of the
transformation $\frac{-4x}{(x^{2}-I_{0})^{2}}$ in the measure, the
probability becomes
\begin{equation}
P=\int_{\sqrt{I_{0}}+\epsilon }^{\lambda }dx\int_{0}^{2k\pi }dv\ \frac{%
4c_{3}^{2}\sin (xv)}{x(x^{2}-I_{0})^{2}}~,~k\in \mathbb{N}.  \label{in1}
\end{equation}%
In order to exclude the case $E=0,\rho =0$ for the initial variables, we
have introduced the constant $\lambda $ as a cut-off. The normalization
gives a quantized value for the constant $c_{3}$. Its qualitative evolution
is given in Fig. \ref{plot111a}.
The qualitative behaviour of the probability function is given by
\begin{equation}
P(x,v)=\int_{\sqrt{I_{0}}+\epsilon }^{x}dx^{\prime }\int_{0}^{v}dv^{\prime
}\ \frac{4c_{3}^{2}\sin (x^{\prime }v^{\prime })}{x^{\prime }(x^{\prime
2}-I_{0})^{2}}  \label{probfun}
\end{equation}%
The surface diagram of this function is presented in Fig. \ref{plot222},
while the contour plot is presented in Fig. \ref{plot333}. The plots show
that for $I_{0}\rightarrow 0$ the probability function reaches its minimum.

At this point, we would like to remind that the Szekeres system admits the
exact solution $u_{A}\left( t\right) =\frac{6^{\frac{2}{3}}}{2}t^{\frac {2}{3%
}},v_{A}\left( t\right) =v_{0}t^{-\frac{1}{3}}$, in which the integration
constants $h$ and $I_{0}$ are zero \cite{anszek}. The latter solution
corresponds to an unstable critical point for the dynamical system %
\eqref{szeksys} and it is very interesting that the conditions for the
existence of the exact solution, i.e. $h=0$ and $I_{0}=0$, lead to an
extremum for the probability function. This might be related with the
existense and the stability of the exact solution. The fact that the quantum
probability has its minimum at the classical value is in accordance with the
analysis of the probability extrema in \cite{dim1} where it was shown that
the extrema of the probability lie on the classical values.

%-----------------------------------------------------------------------------------------------------------------------------------------------------------------------------------------------------------------------------------------------

\section{Conclusions}

%-----------------------------------------------------------------------------------------------------------------------------------------------------------------------------------------------------------------------------------------------
The purpose of this work was the study of the quantum behaviour of the
Szekeres system in the context of the Bohmian interpretation to quantum
theory. The quantization is based on an effective point-like Lagrangian
which can reproduce the two dimensional system of second-order differential
equations resulted from the initial field equations. This Lagrangian is
autonomous, thus there exists a conservation law of \textquotedblleft
energy\textquotedblright\ corresponding to the Hamiltonian function. As for
the extra contact symmetry, it leads to a quadratic in the momenta conserved
quantity attributed to a Killing tensor of the second-rank. The two
conserved quantities give two eigenequations at the quantum level, the
Hamiltonian function being the Schr\"{o}dinger equation.

{The quantum behaviour is studied under the assumption that the wave
function is peaked around its classical value. This leads to the lack of
quantum corrections and the recovery of the classical solutions, thus
leading to the conclusion that the Szekeres universe remains silent at the
quantum level. Finally, for the particular case $h=0$ we study the
probability function and relate one (unstable) exact solution with the
existence of a minimum of this probability. }
%-----------------------------------------------------------------------------------------------------------------------------------------------------------------------------------------------------------------------------------------------

\begin{acknowledgments}
\noindent A.P. acknowledges financial support of FONDECYT grant no. 3160121
and thanks Quatar University for the hospitality provided while part of this
work was carried out. A.Z. acknowledges financial support by the grant GA%
\v{C}R 14-37086G.
\end{acknowledgments}

%-----------------------------------------------------------------------------------------------------------------------------------------------------------------------------------------------------------------------------------------------

%-----------------------------------------------------------------------------------------------------------------------------------------------------------------------------------------------------------------------------------------------

%-----------------------------------------------------------------------------------------------------------------------------------------------------------------------------------------------------------------------------------------------

\end{document}